\documentclass[aps,prd,twocolumn,superscriptaddress,nofootinbib,preprintnumbers]{revtex4}

\usepackage[utf8]{inputenc}

\usepackage{epsfig}
\usepackage{amssymb}
\usepackage{amsmath}
\usepackage{amsfonts}
\usepackage{mathtools}
\usepackage{hyperref}

\newcommand{\be}{\begin{equation}}
\newcommand{\ee}{\end{equation}}
\newcommand{\bea}{\begin{eqnarray}}
\newcommand{\eea}{\end{eqnarray}}
\newcommand{\bi}{\begin{itemize}}
\newcommand{\ei}{\end{itemize}}
\newcommand{\ben}{\begin{enumerate}}
\newcommand{\een}{\end{enumerate}}

\newcommand{\ep}{\epsilon)}

\def\gsim{\mathrel{\rlap{\lower4pt\hbox{\hskip1pt$\sim$}}
    \raise1pt\hbox{$>$}}}         
\def\lsim{\mathrel{\rlap{\lower4pt\hbox{\hskip1pt$\sim$}}
    \raise1pt\hbox{$<$}}}         

\def \ep {\epsilon}

\begin{document}

\title{Quark beam function at next-to-next-to-next-to-leading order in perturbative QCD in the generalized large-$N_c$ approximation}

\author{Arnd Behring}
\email{arnd.behring@kit.edu}
\affiliation{Institute for Theoretical Particle Physics, Karlsruhe Institute of Technology, Karlsruhe, Germany}
\author{Kirill Melnikov}
\email{kirill.melnikov@kit.edu}
\affiliation{Institute for Theoretical Particle Physics, Karlsruhe Institute of Technology, Karlsruhe, Germany}
\author{Robbert Rietkerk}
\email{robbert.rietkerk@kit.edu}
\affiliation{Institute for Theoretical Particle Physics, Karlsruhe Institute of Technology, Karlsruhe, Germany}
\author{Lorenzo Tancredi}
\email{lorenzo.tancredi@physics.ox.ac.uk}
\affiliation{Rudolf Peierls Centre for Theoretical Physics, Clarendon Laboratory, Parks Road, Oxford OX1 3PU, UK}
\author{Christopher Wever}
\email{christopher.wever@tum.de}
\affiliation{Physik-Department T31, Technical University Munich, D-85748, Garching, Germany}

\preprint{TTP19-033, P3H-19-038, TUM-HEP-1234/19, OUTP-19-11P}

\begin{abstract}
We present the matching coefficient for the quark beam function at next-to-next-to-next-to-leading order in perturbative QCD in the generalized large $N_c$-approximation, $N_c \sim N_f \gg 1$. Although several refinements are still needed to make this result interesting for phenomenological applications, our computation shows that a fully-differential description of simple color singlet production processes at a hadron collider at N$^3$LO in perturbative QCD  is within reach.
\end{abstract}

\maketitle

\allowdisplaybreaks

\section{Introduction}
Good  understanding of infra-red and collinear limits in perturbative QCD and the ability to use
this understanding for an increasingly accurate description of  hadron collisions  is one of the
key  elements for the success of the future LHC physics program.  Because of that,
much of the  current effort  in theoretical collider physics
focuses  on achieving and advancing such  understanding  in a number  of complementary ways, 
ranging  from  fixed-order computations,  to resummations and, finally, to parton showers.
Although for each of these approaches there exists a  set of  observables and theoretical quantities to
which it is traditionally applied,  there
are a few cases which lie at  their  intersections and where progress achieved in the context of one approach
has implications for the other ones. 

One such theoretical quantity  is the so-called beam function \cite{Stewart:2009yx,Stewart:2010qs}.  Beam functions describe
the  dynamics of incoming partons that slightly
deviate from their original direction by emitting hard quasi-collinear radiation before going into the hard process.
For this reason, beam functions  are  important ingredients for  resummation studies that aim to understand differential
cross sections in the  quasi-collinear region \cite{Becher:2012yn,Lustermans:2019plv,Ebert:2016gcn,Chen:2018pzu,Bizon:2019zgf}.

Two  collinearity measures have been discussed  in the literature -- 
the total transverse momentum of the radiated partons
\be
p_\perp = \left|\sum \limits_{j=1}^{N} \vec k_{j,\perp}\right|, 
\label{eq0a}
\ee
and the 0-jettiness 
\be
{\cal T} = \sum \limits_{j=1}^{n} \min_{i \in \{1,2\}} \left [ \frac{2 p_i \cdot k_j}{Q_i} \right ].
\label{eq0}
\ee
In Eqs.~(\ref{eq0a})--(\ref{eq0}), $Q_{1,2}$ are so-called ``hardness'' variables for the initial state partons 
(see, e.g., \cite{Stewart:2009yx,Billis:2019vxg}), $p_{1,2}$ are the momenta of the incoming partons and $k_{1,\dots,N}$ are the momenta
of on-shell final state  partons. 

As was shown in Refs.~\cite{Stewart:2009yx,Stewart:2010qs} using soft-collinear effective field theory (SCET) \cite{Bauer:2000ew,Bauer:2000yr,Bauer:2001ct,Bauer:2001yt,Bauer:2002nz},
beam functions are non-perturbative objects that can be perturbatively matched to parton distribution functions in case a collinearity measure exceeds $\Lambda_{\rm QCD}$.
Perturbative matching coefficients can then be used to construct slicing schemes for higher-order computations as proposed in Refs.~\cite{Catani:2007vq,Grazzini:2008tf,Boughezal:2015dva,Gaunt:2015pea}.   
Currently, all matching coefficients for both $p_\perp$ and 0-jettiness beam functions are known through next-to-next-to-leading order (NNLO) in QCD \cite{Gehrmann:2012ze,Gaunt:2014xga,Gaunt:2014cfa,Boughezal:2017tdd}.

It is quite  interesting to extend the  computation of the matching coefficients 
to one order higher in the strong coupling constant $\alpha_s$.  Not only will such a computation   stress-test many aspects
of our understanding of soft-collinear dynamics in QCD, as well as many techniques of
perturbative quantum field theory, but it will also provide an alternative path to next-to-next-to-next-to-leading order (N$^3$LO) QCD description of 
color-singlet production at the {\it exclusive} level. Currently,  N$^3$LO QCD corrections to the {\it inclusive}
 cross section \cite{Anastasiou:2015ema,Anastasiou:2016cez,Dulat:2017prg,Mistlberger:2018etf,Dulat:2018rbf},
as well as to the Higgs rapidity distribution  in Higgs boson production in gluon fusion  are available
\cite{Dulat:2018bfe}. An extension of N$^3$LO  computations to Drell-Yan-like processes, accounting for  decays of
$Z$ and $W$ bosons  to leptons, is very desirable.

   Recently, we have computed  loop and phase-space  integrals relevant for the 
 so-called triple-real and double-real single-virtual contributions to the quark-to-quark 0-jettiness 
 matching coefficient, focusing on gluonic final states \cite{Melnikov:2018jxb,Melnikov:2019pdm}.
 When combined with the computation of the single-real double-virtual splitting
 function $q^* \to qg $ described in Ref.~\cite{Duhr:2014nda}, all  ingredients required 
to  obtain  the N$^3$LO QCD contribution to the quark-to-quark matching coefficient ${\cal I}_{qq}$
{\it through leading color} become available. In addition, the results reported in
\cite{Melnikov:2018jxb,Melnikov:2019pdm} allow us to compute all N$^3$LO contributions that scale as $N_c^2 N_f$ and
$N_c N_f^2$, where $N_f$ is the number of massless quarks in the theory.

The goal of this paper  is to present
the N$^3$LO contribution to the quark matching coefficient in the approximation $N_c \sim N_f \gg 1$, keeping
only leading  ${\cal O}(\alpha_s^3 N_c^3, \alpha_s^3 N_f N_c^2, \alpha_s^3 N_f^2 N_c)$ terms.
We will refer to it  as the generalized large-$N_c$ or leading-color  approximation.

We note that our computation of the matching coefficient  ${\cal I}_{qq}$  is restricted to generalized leading-color approximation
since, so far, we have not computed 
all the required contributions of  final states with additional quark pairs that are relevant  beyond the generalized large-$N_c$ limit. In principle,
the required computations are similar to what has already been
done in Refs.~\cite{Melnikov:2018jxb,Melnikov:2019pdm} but, due to proliferation 
of integrals  required for multi-quark final states, the calculations
have  not been finalized.

Nevertheless, we   believe that the generalized  large-$N_c$ N$^3$LO contribution to the quark-to-quark
matching coefficient  is an interesting  intermediate result
since, at variance with our previous publications~\cite{Melnikov:2018jxb,Melnikov:2019pdm}, it explicitly demonstrates how different pieces combine 
to produce  a well-defined physical quantity   at next-to-next-to-next-to-leading order in perturbative QCD. 
 It also shows that such high-order computations, in spite of their significant complexity, appear to be
 doable with current computational technologies. 

The rest of the paper is organized as follows. In  Section~\ref{sect2} we describe how the computation of the
perturbative matching coefficient  is set up. In Section~\ref{sect3}
 we discuss how the various  required ingredients are obtained. 
 We present the result for the matching coefficient in the generalized large-$N_c$ approximation 
 in Section~\ref{sect4}  and conclude in Section~\ref{sect5}. A number of useful formulas can be found in 
 the Appendix.

 \section{Perturbative matching coefficient}
 \label{sect2}

In this section we explain how the perturbative matching coefficient is computed. The starting
point is the relation between beam functions and parton distribution functions
\begin{align}
\widetilde{B}_{i}(t,z,\mu) 
= \sum_{k} \mathcal{I}_{ik}(t,z,\mu) \underset{z}{\otimes} \widetilde{f}_{k}(z,\mu) \,, 
\label{eq:beam_function_non_perturbative}
\end{align}
where the sign $\underset{z}{\otimes}$ stands for the convolution\footnote{We have used the program
\texttt{MT} \cite{Hoeschele:2013gga} to compute the $z$-convolutions required for the matching coefficient computation.}
\begin{align} 
f(z) \underset{z}{\otimes} g(z) 
 = \int\limits_0^1 dz_1 dz_2 f(z_1) g(z_2) \delta(z-z_1 z_2).
\end{align}
The proportionality coefficients between the beam functions and the parton distribution functions, 
$\mathcal{I}_{ik}(t,z,\mu)$
in Eq.~(\ref{eq:beam_function_non_perturbative}), 
are the matching coefficients. The sum in Eq.~(\ref{eq:beam_function_non_perturbative}) runs over all species 
of partons that are found in the proton
for a particular value of
the factorization scale $\mu$.  The parameter $t$ is the so-called transverse virtuality, which  is related to the 0-jettiness
variable ${\cal T}$ in Eq.~(\ref{eq0}) and will be defined below in Eq.~(\ref{eq12}).

For $\sqrt{t} \gg \Lambda_{\rm QCD}$, the matching coefficient $\mathcal{I}_{ik}$ can be calculated in perturbative QCD. To this
end,   we replace the non-perturbative parton distributions with their
perturbative counter-parts, calculate the partonic beam function and extract the matching coefficient 
by comparing the two sides
of Eq.~(\ref{eq:beam_function_non_perturbative}). Similar to parton distribution functions, this can be done for
any combination of an incoming parton $j$ and the parton $i$ that eventually goes into the hard scattering. We therefore write 
\begin{align}
B_{ij}(t,z,\mu) = \sum_{k \in \{q,\bar{q},g\}} \mathcal{I}_{ik}(t,z,\mu) \underset{z}{\otimes} f_{kj}(z,\mu).
\label{eq:beam_function}
\end{align}
In contrast to Eq.~(\ref{eq:beam_function_non_perturbative}), all quantities in  Eq.~(\ref{eq:beam_function}) admit 
an  expansion in the strong coupling constant $\alpha_s$. Writing 
\be
\begin{split} 
  & B_{ij} = \sum \limits_{n=0}^{\infty} \left ( \frac{\alpha_s}{4 \pi} \right )^n \; B_{ij}^{(n)},\;\;\; \\
  & \mathcal{I}_{ij} = \sum \limits_{n=0}^{\infty} \left ( \frac{\alpha_s}{4 \pi} \right )^n  \; \mathcal{I}_{ij}^{(n)}, \\
  & f_{ij} = \sum \limits_{n=0}^{\infty} \left ( \frac{\alpha_s}{2 \pi} \right )^n  \; f_{ij}^{(n)},
\end{split} 
\label{eq4}
\ee
and defining the leading-order quantities through $B_{ij}^{(0)} = \delta_{ij} \delta(t) \delta(1-z)$, 
${\cal I}_{ij}^{(0)} = \delta_{ij} \delta(t) \delta(1-z)$ and $f_{ij}^{(0)} = \delta_{ij} \delta(1-z)$, 
we solve Eq.~(\ref{eq:beam_function}) to express the matching coefficients through the 
partonic  beam function.
We find 
\be
\begin{split} 
\mathcal{I}_{ij}^{(1)}(t,z,\mu) = {}&
B_{ij}^{(1)}(t,z,\mu) 
- 2 \delta(t) f_{ij}^{(1)}(z)
\,,\\
\mathcal{I}_{ij}^{(2)}(t,z,\mu) = {}&
B_{ij}^{(2)}(t,z,\mu) 
- 4 \delta(t) f_{ij}^{(2)}(z) 
\\
& - 2 \sum_k \mathcal{I}_{ik}^{(1)}(t,z,\mu) \underset{z}{\otimes} f_{kj}^{(1)}(z)
\,,\\
\mathcal{I}_{ij}^{(3)}(t,z,\mu) = {}&
B_{ij}^{(3)}(t,z,\mu) 
- 8 \delta(t) f_{ij}^{(3)}(z) 
\\
&
- 4 \sum_k \mathcal{I}_{ik}^{(1)}(t,z,\mu) \underset{z}{\otimes} f_{kj}^{(2)}(z)
\\
& - 2 \sum_k \mathcal{I}_{ik}^{(2)}(t,z,\mu) \underset{z}{\otimes} f_{kj}^{(1)}(z).
\label{eq:I_3}
\end{split}
\ee

Perturbative parton distribution functions in various orders in $\alpha_s$
are obtained as iterative solutions of the Altarelli-Parisi equation
\begin{align}
\mu^2 \frac{d}{d\mu^2} f_{ij}(z) = \frac{\alpha_s}{2 \pi} \sum_{k} P_{ik}(z) \underset{z}{\otimes} f_{kj}(z)\,,
\label{eq:DGLAP}
\end{align}
with the boundary condition given above.  We note that  since in Eq.~(\ref{eq:beam_function_non_perturbative}) 
the parton distribution
functions are the ${\overline {\rm MS}}$ ones,  the perturbative parton distribution functions that we need
can only contain poles in the dimensional regularization parameter $\epsilon$.
Explicit  results for $f_{ij}^{(1,2,3)}$ in terms of the splitting functions $P_{ik}$ 
are given  in the Appendix. 

Eq.~(\ref{eq:I_3}) allows us to iteratively compute the matching coefficients once the perturbative beam functions
become available. 
However, a beam function computed directly from the quasi-collinear limits of the relevant scattering amplitudes 
is what one refers to as a {\it bare} beam function, because it contains both soft and collinear divergences. 
Soft divergences must be  removed 
by a dedicated $\overline {\rm MS}$-subtraction that, schematically,  is given by the following formula \cite{Stewart:2009yx}
\be
B_{ij}^{\rm b}(t,z) = Z_{i}(t,\mu) \underset{t}{\otimes} B_{ij}(t,z,\mu).
\label{eq8}
\ee
In Eq.~(\ref{eq8})  the convolution  with respect to  $t$ is defined by the equation 
\be
f(t) \underset{t}{\otimes} g(t) 
= \int\limits_0^\infty dt_1 dt_2 f(t_1) g(t_2) \delta(t-t_1-t_2).
\ee

To compute the quark-to-quark matching coefficient, we require the renormalization constant $Z_q$. 
Similar to other renormalization constants, $Z_q$ satisfies a 
renormalization group equation \cite{Stewart:2009yx}
\be
\mu \frac{{\rm d}}{{\rm d} \mu } Z_q(t,\mu) = - Z_q(t,\mu) \underset{t}{\otimes} \gamma_q(t,\mu),
\label{eq9}
\ee
where the anomalous dimension reads
\be
\gamma_q(t,\mu) = \gamma_B^{q}\delta(t) - 2 \Gamma_{\rm cusp}^{q}  L_0 \left (\frac{t}{\mu^2} \right ).
\label{eq10}
\ee
The anomalous dimensions $\gamma_B^{q}$ and $\Gamma^{q}_{\rm cusp}$ are known through  ${\cal O}(\alpha_s^3)$ \cite{Stewart:2010qs,Korchemsky:1987wg,Moch:2004pa,Vogt:2004mw}.
Here, $L_0(t/\mu^2)$ is the modified plus-distribution $L_0(t/\mu^2) = \mu^{-2} [\mu^2/t]_+$ with the (regularized) singularity at $t/\mu^2 = 0$ rather than at $t/\mu^2=1$. 
In practice, we construct the renormalization constant $Z_q$ in the $\overline {\rm MS}$-scheme from Eq.~(\ref{eq9}) by expanding the various quantities in the strong coupling constant, see e.g. Eq.~(\ref{eq:Z_inverse_expansion}), and inserting an ansatz for $Z_q$ in terms of $t$-distributions. The ansatz is constructed following an observation that $Z_q$ must have the same $t$-dependence as the bare beam function in order to cancel the soft divergences.
We then use Eq.~(\ref{eq8}) to obtain the renormalized partonic beam function from the bare one. 
Finally, we employ Eq.~(\ref{eq:I_3}) to derive the desired matching coefficient.  
Explicit formulas for various steps described above are given in the Appendix. 

We note that since the partonic PDFs are singular in the $\ep \to 0$ limit, $f_{ij}^{(n)} \sim \ep^{-n}$,  it follows from  Eq.~(\ref{eq:I_3}) that the matching coefficients  ${\cal I}_{ij}^{(1,2)}$ need to be known to higher powers  in  the dimensional regularization parameter $\ep$. 
The relevant computation was performed in Ref.~\cite{daniel} and we borrow the  results from there. 

It remains to discuss the computation of the {\it bare} beam function. We do that in the next  section.

\section{Computation of the bare 0-jettiness quark  beam function}
\label{sect3}

It is clear  that the  major challenge for  computing matching coefficients through  third order in
perturbative QCD is  the calculation  of the bare beam functions.  We can obtain the   bare 
quark beam function from any physical process that features
a quark in the initial state,  by   extracting the
leading collinear-enhanced contributions. Since leading collinear singularities factorize into products
of universal splitting functions 
and hard matrix elements, one can organize the calculation in a process-independent way. 

Indeed, in physical gauges,   collinear splitting functions  can be obtained by considering 
QCD radiation {\it off a single external line}~\cite{Catani:1999ss}, for example  the incoming quark line in our case.
It is important that the emissions, both real and virtual, 
that originate  from  any other incoming lines, do not contribute to leading collinear singularities and, for this reason, 
can be ignored.   The splitting functions so obtained must be
integrated over the particular phase space for real  emission(s)
that is constrained in such a way as to keep the momentum fraction $z$ and the transverse virtuality $t$ of the incoming quark that goes into the hard scattering process fixed~\cite{Ritzmann:2014mka}.

The bare quark beam function at N$^3$LO is then computed 
by adding such collinear-enhanced  contributions with up to three real partons  in the final state, with 
the
number of virtual loops required to provide the ${\cal O}(\alpha_s^3)$ correction to the leading-order transition $q \to q$. 
Hence, we need to consider a  tree-level contribution where a quark splits into a virtual quark that goes into a hard process 
and three real partons, a one-loop correction  to a process where a quark splits into a virtual quark and two real partons and
a two-loop correction to the $q \to q^* + g$   splitting.

\begin{figure}
\centering
\includegraphics[width=0.47\textwidth]{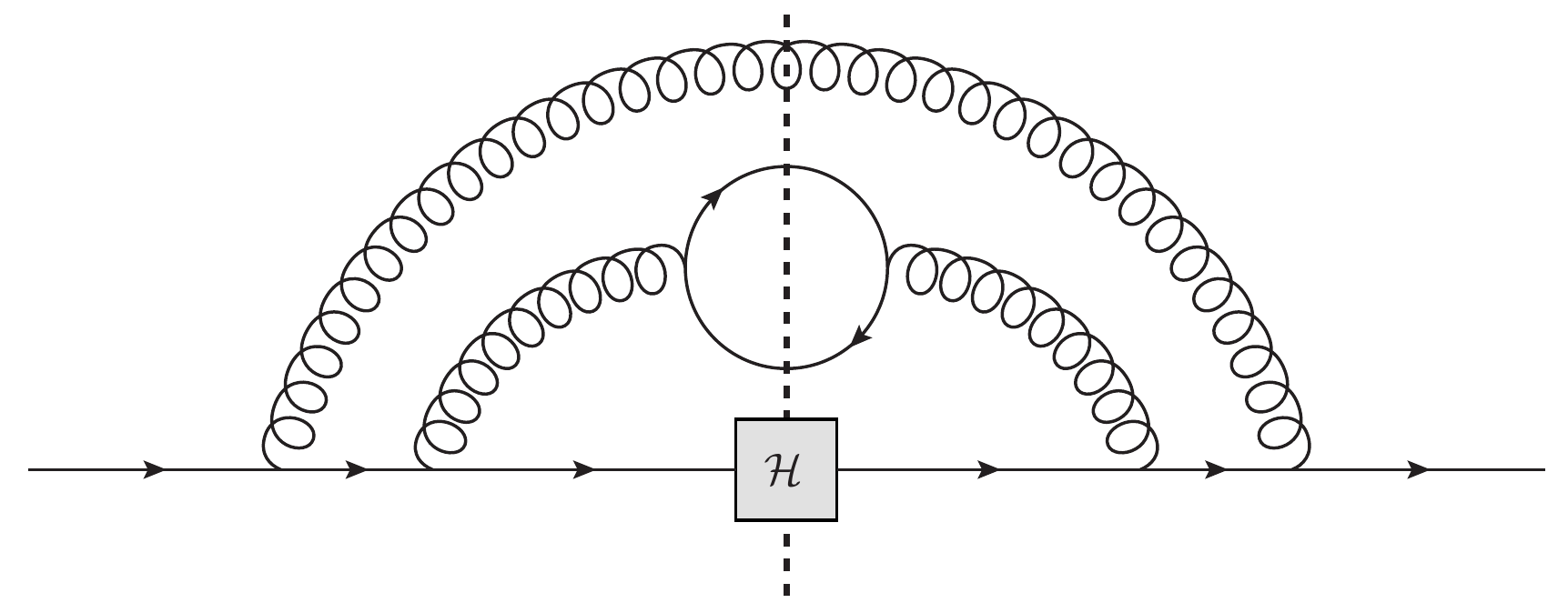}\\[0.5em]
\includegraphics[width=0.47\textwidth]{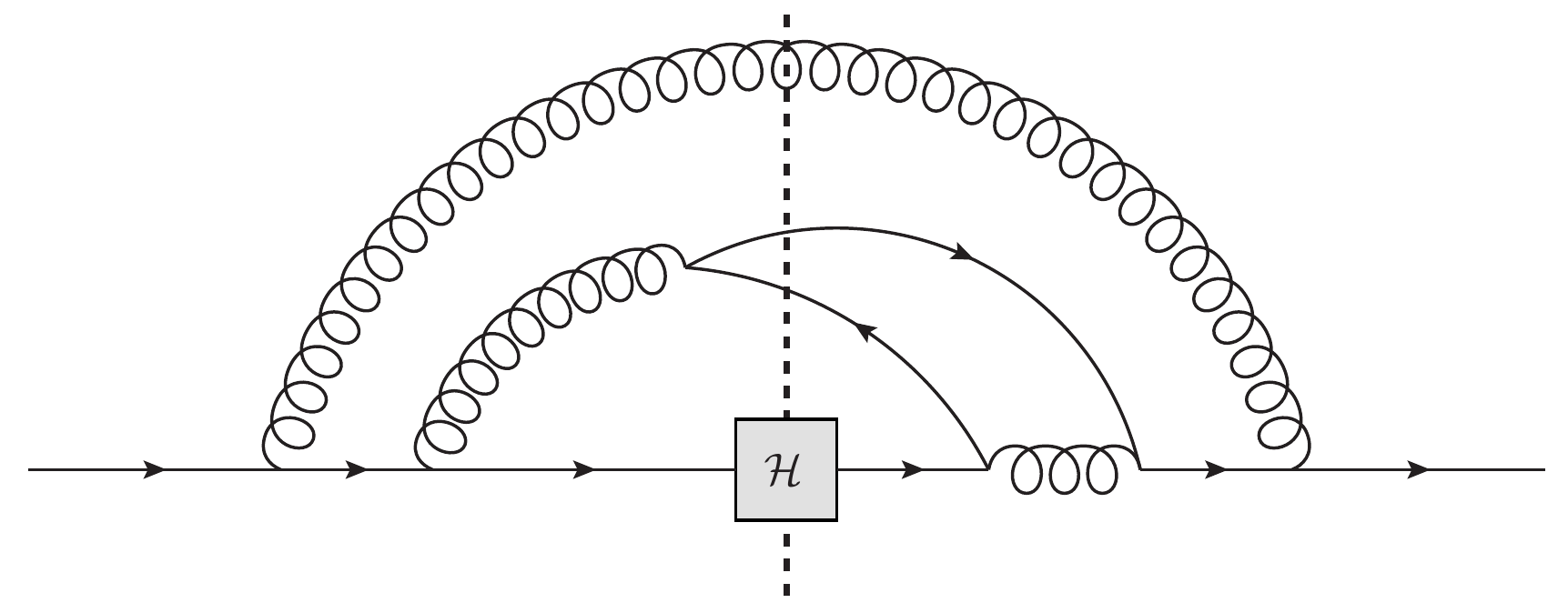}\\[-0.5em]
\caption{
Top: example of a triple-real emission amplitude with a quark-antiquark pair in the final state which contributes to the bare beam function in the leading-color approximation and therefore has been included in our computation.
Bottom: example of a similar amplitude which is sub-leading in $N_c$ and therefore is not included in our computation.
The box labeled $\mathcal{H}$ denotes the hard scattering process.
\vspace{-4mm}
}
\label{fig:example_LC_NLC}
\end{figure}

Since in this paper we focus on the generalized large-$N_c$ contribution to the quark beam function, where the number of colors and the number of flavors are taken to be large $N_c \sim N_f \gg 1$, it is sufficient to consider gluons in the final state as well as quarks that exclusively originate from a final-state gluon splitting. 
Other final states are sub-leading in the generalized large-$N_c$ approximation. 
Fig.~\ref{fig:example_LC_NLC} illustrates which types of quark-antiquark final states have been included and which types have been excluded from our calculation.

We schematically write the ${\cal O}(\alpha_s^3)$  contribution to the bare   beam function of a quark 
in the following way 
\be
B_{qq}^{{\rm b},(3)} =
B_{qq}^{{\rm b},(R3V0)} + B_{qq}^{{\rm b},(R2V1)} + B_{qq}^{{\rm b},(R1V2)}\,,
\label{eq131}
\ee
where the label $Rn_RVn_V$ refers to processes 
with $n_R$ real partons  and  $n_V$ virtual loops.  The quantities $B_{qq}^{{\rm b},Rn_RVn_V}$ read  
\be
\begin{split} 
& B_{qq}^{{\rm b},Rn_RVn_V}(t,z) \sim
\int \prod \limits_{i=1}^{n_R}  [{\rm d} k_i]  \delta \left ( 2 p \cdot k_{n_R} - \frac{t}{z}
\right )
\\
& \times \delta \left ( \frac{2 \bar p \cdot k_{n_R}}{s} - (1-z) \right ) \; P^{(Rn_RVn_V)}_{qq}(p, \bar p, \{k_i\}), 
\end{split}
\label{eq12}
\ee
where 
$p$ is the four-momentum of the incoming parton, $\bar p$ is the complementary  collinear direction, $s = 2 p \cdot \bar p$, 
$
[dk_i] = {\rm d}^{d-1} k_i/( (2\pi)^{d-1} 2k_i^{(0)})
$
is a single-parton phase-space element, 
$k_{n_R} = \sum \limits_{i=1}^{n_R} k_i$ and $P_{qq}^{(Rn_RVn_V)}$ denotes  the $n_V$-loop contribution
to the collinear splitting functions that describes the $q \to q^* + g_1+...+g_{n_R}$ process or, if $n_R \geq 2$, the $q \to q^* + q' + \bar{q}' + g_3+...+g_{n_R}$ process.  We note that
the functions $B_{qq}^{{\rm b},Rn_RVn_V}(t,z)$ scale  uniformly  with the transverse virtuality, i.e. 
\be
B_{qq}^{{\rm b},Rn_RVn_V}(t,z) \sim t^{-1-3\ep} {\tilde B}_{qq}^{{\rm b},Rn_RVn_V}(z) .
\label{eq15}
\ee
This observation will be important for the discussion below  where we describe the computation  of  the double-virtual
single-real contribution $B_{qq}^{{\rm b},R1V2}$.

The calculation of the triple-real and double-real single-virtual contributions  $B_{qq}^{{\rm b},R3V0}$ and $B_{qq}^{{\rm b},R2V1}$ 
was discussed   in 
Refs.~\cite{Melnikov:2018jxb,Melnikov:2019pdm}, respectively.
We will briefly summarize these  discussions here.

Although, as we already said, the 
collinear splitting functions in Eq.~(\ref{eq12}) are universal objects, they are not available
in closed form beyond NNLO. Since, as shown in Eq.~(\ref{eq12}), our goal is not only to construct 
the splitting functions, but also to  integrate
them over the real-emission phase space, it is important to have an algorithm that allows us to perform  both of these
tasks in a concerted  way. We achieve this by following the procedure outlined in Ref.~\cite{Catani:1999ss}
that describes how to extract splitting functions by considering  emissions off a single external line and by
employing
relevant  projection operators. An important
ingredient in this construction is the use of physical gauges for both virtual and real gluons that, unfortunately, 
 complicates the computations significantly. In Ref.~\cite{Catani:1999ss} this procedure was used to
explicitly construct all tree-level splitting functions at NNLO in QCD. 
Here, we just use this procedure to find
a suitable expression for the  collinear splitting functions $P^{(Rn_RVn_V)}_{qq}(p, \bar p, \{k_i\})$ that may involve
unintegrated momenta of both real and virtual gluons. 
Once such a representation for $P^{(Rn_RVn_V)}_{qq}(p, \bar p, \{k_i\})$ is available, we 
apply reverse unitarity \cite{Anastasiou:2002yz} to map phase-space  integrals  onto loop integrals.
We then use   integration-by-parts technology \cite{Tkachov:1981wb,Chetyrkin:1981qh} to express each particular
contribution to $B_{qq}^{\rm bare}$ in terms of master integrals and to derive the differential equations that these
integrals satisfy \cite{Kotikov:1990kg,Bern:1993kr,Remiddi:1997ny,Gehrmann:1999as}.

A detailed discussion of how the master integrals are computed  from the
relevant differential equations can be found in Refs.~\cite{Melnikov:2018jxb,Melnikov:2019pdm}.  Here, we just note that
the use of physical gauges makes their computation much more difficult, in that it introduces additional propagator-like
structures that arise from polarization sums of real and virtual gluons.  Unfortunately, this leads to a proliferation
of  integrals that need to be calculated. 
Another interesting point is that the master integrals, that describe triple-real emissions, are initially written as linear combinations of  generalized polylogarithms of a complex-valued variable
\be
x = -1 +\frac{z}{2} \pm \frac{i}{2} \sqrt{z (4-z)},
\ee
which arises during the rationalization of the differential equations, see Ref.~\cite{Melnikov:2019pdm}.
Curiously, as we will see from the final result, the dependence on $x$ disappears once the 
complete  triple-real emission contribution
to the beam function is constructed. 

In principle, one can compute the $B_{qq}^{{\rm b},R1V2}$ contribution to the beam function using a similar approach. This
would require the calculation of  the two-loop correction to the process $q \to q^* + g$ in a physical gauge; such
computation is, currently, not available. 
Fortunately, there is a way out.
The contribution  we are interested in can be extracted from the two-loop amplitude of 
the  process $q(p)  \bar q(\bar p)  \to V+g(k_1) $ in the limit when the gluon is emitted along the direction
of the incoming quark $q$.   To see this, consider the Mandelstam variables  $T = (p-k_1)^2$, $U = (\bar p - k_1)^2$ and $S= 2p \cdot  \bar p$
that are needed to describe this process. 
Then, from the phase-space constraints in Eq.~(\ref{eq12}),
we find  $T = -t/z,\;U = -s(1-z)$. 
Therefore, we can obtain the required splitting function by studying the $T \to 0$ limit of the NNLO QCD contribution to the amplitude squared for the process $q(p)  \bar q(\bar p)  \to V+g(k_1)$, and by extracting the contribution with the appropriate $T^{-1-2\ep}$ scaling.\footnote{According to Eq.~(\ref{eq15}), the N$^3$LO contributions to
the beam functions scale as $t^{-1-3\ep}$. In case of the double-virtual single-real term $B_{qq}^{{\rm b},R1V2}$, this scaling is obtained from the $t^{-1-2\ep}$ scaling of the virtual amplitude squared and the $t^{-\ep}$ scaling of the single-gluon phase space.}
The calculation of the $0 \to q \bar q  Vg$ scattering amplitude in the $T \to 0$ limit is  available \cite{Duhr:2014nda},
so that the splitting function $P^{(R1V2)}_{qq}$ can be extracted from that reference.  
An analytic continuation is required to obtain the initial-state splitting function from the final-state one; this can be done following the discussion in Ref.~\cite{Duhr:2014nda}. 
For the correct regularisation of the soft limit $z \to 1$ it is important to keep also factors of
$(1-z)^{-a\ep}$ unexpanded in $\ep$, which fortunately is the case in that reference. Finally, we note that
the remaining  integration over the single-gluon phase space is straightforward  since the  phase-space  constraints restrict the
gluon kinematics to a point that, in fact, no non-trivial integration is needed. 
The integration over the singular limits of the
single-real emission phase space introduces up to two additional powers of $\ep^{-1}$ so that, in order to correctly obtain the
$\ep^0$ term of the bare beam function, the first six orders of the expansion in $\ep$ of the splitting function have to be known.
Ref.~\cite{Duhr:2014nda} contains the first five orders of the splitting function, but the sixth order is only necessary for
the soft limit $z \to 1$, so that it can be reconstructed from the soft current calculated in Ref.~\cite{Duhr:2013msa}, see Ref.~\cite{Duhr:2014nda} for more details. 

In addition to the two-loop virtual corrections to the $q \to q^* + g$ process, 
the square of the one-loop correction to the single-gluon emission process has to be included into the calculation of $B_{qq}^{{\rm b},R1V2}$.  We  obtained this contribution  
by adapting  the computation of the NNLO QCD bare beam function to higher orders in dimensional regularization
parameter $\ep$, as reported in Ref.~\cite{daniel}.

\section{Result for the matching coefficient}
\label{sect4}

We are now in a position to present  the N$^3$LO   contribution to the quark
matching coefficient in the generalized large-$N_c$ approximation.  To this end,  we write the  ${\cal O}(\alpha_s^n)$ contribution
to the matching coefficient, as defined in Eq.~(\ref{eq4}), in the following way
\be
   {\cal I}^{(n)}_{qq} = \sum \limits_{k=0}^{2n-1} L_{k}\left(\frac{t}{\mu^2} \right ) \; F^{(n,k)}_{+}(z) 
   + \delta(t) F_{\delta}^{(n)}(z)\,,
\ee
where $L_k(t/\mu^2) = 1/\mu^2[\ln^k(t/\mu^2)/(t/\mu^2)]_+$.
Furthermore,  it is useful to isolate the so-called soft contributions in $F_{\delta}^{(n)}(z)$. These contributions 
contain $\delta(1-z)$ and the plus-distributions $D_k(z) = [\ln^k(1-z)/(1-z)]_+$; all other terms in $F_{\delta}^{(n)}(z)$ are
referred to as ``hard''.
We therefore write 
\be
F_{\delta}^{(n)}(z)= C^{(n)}_{-1} \delta(1-z) + \sum \limits_{k=0}^{2n-1} C^{(n)}_{k} D_k(z) + F_{\delta,{\rm h}}^{(n)}(z).
\ee
As we already mentioned, the NLO and NNLO contributions to the matching coefficient ${\cal I}_{qq}^{(1),(2)}$ are fully
known \cite{Gaunt:2014xga,Boughezal:2017tdd}.
Recently, in Ref.~\cite{Billis:2019vxg}, it was shown how to extract the
soft contributions to N$^3$LO matching coefficient described by  the constants $C^{(3)}_{k}$, $k=-1,\dots,5$ from known results
in the literature \cite{Lustermans:2019cau,Ahmed:2014uya,Dulat:2018bfe,Li:2016ctv,Ravindran:2006bu,Li:2016axz}. Also, by using the
renormalization group equations for the matching coefficient, all functions $F^{(3,k)}_{+}(z)$ were calculated in that
reference. These results, especially the ones for the soft constants, provide 
an important check on the correctness of  our  computation. Indeed, we have verified that our results 
reproduce the constants $C^{(3)}_{k}$, $k = -1,\dots,5$
and the functions $F_+^{(3,k)}(z)$  reported in  Ref.~\cite{Billis:2019vxg} in the limit $N_c \sim N_f \gg 1$.

The new result of this paper is the contribution of hard collinear gluons
to the function $F_{\delta}^{(3)}(z)$ in the generalized large-$N_c$ limit. 
The result turns out to be remarkably simple. It is expressed 
in terms of harmonic polylogarithms of the variable $z$ of up to weight five.
To present the result in a compact form, we use a notation for harmonic polylogarithms (HPLs) introduced
in Ref.~\cite{Remiddi:1999ew} and extended in Ref.~\cite{Maitre:2005uu}. To this end,   we explicitly list the right-most zeros of an HPL  index but, starting
from the first non-vanishing entry,  we do not display trailing zeros in an index anymore.  Instead, we  add one to the absolute value
of the index  entry per trailing zero and continue doing so until the next non-zero entry is reached. 
For example, in the formulas below, $H_{1,2,1,0}$ means $H(1,0,1,1,0,z)$ whereas   $H_{4,1}$ is $ H(0,0,0,1,1,z)$ etc.
Armed with this understanding, we present the result for the hard contribution to ${\cal I}^{(3)}_{qq}(t,z)$ in the generalized large-$N_c$
approximation.
To this end, we write
\begin{align}
F^{(3)}_{\delta,{\rm h}} = N_f^2 N_c  T_R^2 F_1 +  N_f N_c^2  T_R F_2 + N_c^3F_3,
\end{align}
where $T_R = 1/2$. We note that all other contributions are subleading either in $N_f$ or in $N_c$ and are thus neglected.  
The three functions read
\begin{widetext}
\begin{align}
F_{1}(z) ={}&
        \frac{32}{729} (157 z-41)
        +\frac{80}{81} (11 z-1) H_1
        +\frac{64}{27} (4 z+1) H_{1,1}
        +\frac{32}{9} (z+1) H_{1,1,1}
        -\frac{16}{27} (z+1) \pi ^2 H_1 
\nonumber \\&
        +\frac{1}{1-z}\bigg[
                -\frac{32}{81} \left(49 z^2-32 z+34\right) H_0
        \bigg]
        +\frac{1}{1-z}\bigg[
                -\frac{32}{27} \left(16 z^2-9 z+13\right) H_2
                -\frac{64}{27} \left(4 z^2-3 z+4\right) H_{1,0}
\nonumber \\&
                -\frac{16}{81} \left(133 z^2-60 z+97\right) H_{0,0}
                +\frac{16}{81} \pi ^2 \left(16 z^2-9 z+3\right)
        \bigg]
        +\frac{1}{1-z}\bigg[
                -\frac{32}{3} \left(z^2+1\right) H_3 
                -\frac{64}{9} \left(z^2+1\right) H_{2,1}
\label{eq19}
\\&
                -\frac{64}{9} \left(z^2+1\right) H_{2,0}
                -\frac{32}{9} \left(z^2+1\right) H_{1,2}
                -\frac{32}{9} \left(z^2+1\right) H_{1,1,0}
                -\frac{32}{9} \left(z^2+1\right) H_{1,0,0}
                -\frac{368}{27} \left(z^2+1\right) H_{0,0,0}
\nonumber \\&
                +\frac{16}{9} \left(z^2+1\right) \pi ^2 H_0 
                +\frac{64}{27} \left(z^2+2\right) \zeta _3 
        \bigg]
,
\nonumber \\[1em]
F_{2}(z) ={}&
        \frac{1}{2916}(96373-401039 z)
        +\frac{1}{162} (2075-21433 z) H_1 
        -\frac{2}{27} (1301 z+215) H_{1,1}
        -\frac{8}{9} (67 z+37) H_{1,1,1}
\nonumber \\&
        -\frac{80}{3} (z+1) H_{1,1,1,1}
        +\frac{1}{1-z}\bigg[
                \frac{1}{162}\left(33155 z^2-25816 z+27301\right) H_0
        \bigg]
\nonumber \\&
        +\frac{1}{1-z}\bigg[
                \frac{2}{81} \left(6683 z^2-4254 z+5375\right) H_2
                +\frac{2}{81} \left(3845 z^2-3048 z+3917\right) H_{1,0}
\nonumber \\&
                +\frac{1}{243} \pi ^2 \left(-6389 z^2+3606 z-307\right)
        \bigg]
        +\frac{1}{1-z}\bigg[
                \frac{4}{9} \left(273 z^2-73 z+209\right) H_3
                +\frac{4}{9} \left(206 z^2-83 z+185\right) H_{2,1}
\nonumber \\&
                +\frac{4}{27} \left(521 z^2-168 z+461\right) H_{2,0}
                +\frac{4}{9} \left(157 z^2-63 z+164\right) H_{1,2}
                +\frac{4}{9} \left(117 z^2-49 z+110\right) H_{1,1,0}
\nonumber \\&
                +\frac{8}{27} \left(176 z^2-54 z+185\right) H_{1,0,0}
                +\frac{2}{27} \left(1477 z^2-249 z+922\right) H_{0,0,0}
\label{eq20} \\&
                +\frac{1}{27} \left(-387 z^2+162 z+65\right) \pi ^2 H_1
                -\frac{2}{27} \left(319 z^2-94 z+234\right) \pi ^2 H_0
                -\frac{4}{9} \left(225 z^2-76 z-108\right) \zeta _3
        \bigg]
\nonumber \\&
        +\frac{1}{1-z}\bigg[
                \frac{2}{9} \left(319 z^2+12 z+193\right) H_4
                +\frac{8}{3} \left(24 z^2+17\right) H_{3,1}
                +\frac{8}{9} \left(62 z^2-3 z+38\right) H_{3,0}
\nonumber \\&
                +\frac{4}{3} \left(31 z^2+2 z+27\right) H_{2,2}
                +\frac{4}{9} \left(91 z^2+73\right) H_{2,1,1}
                +\frac{4}{9} \left(79 z^2-6 z+55\right) H_{2,1,0}
                +\frac{8}{9} \left(29 z^2+20\right) H_{2,0,0}
\nonumber \\&
                +\frac{356}{9} \left(z^2+1\right) H_{1,3}
                +\frac{344}{9} \left(z^2+1\right) H_{1,2,1}
                +\frac{248}{9} \left(z^2+1\right) H_{1,2,0}
                +\frac{232}{9} \left(z^2+1\right) H_{1,1,2}
\nonumber \\&
                +\frac{64}{3} \left(z^2+1\right) H_{1,1,1,0}
                +\frac{116}{9} \left(z^2+1\right) H_{1,1,0,0}
                +\frac{172}{9} \left(z^2+1\right) H_{1,0,0,0}
                +\frac{2}{9} \left(201 z^2+19\right) H_{0,0,0,0}
\nonumber \\&
                -\frac{2}{9} \left(41 z^2+2 z+37\right) \pi ^2 H_2
                -\frac{2}{27} \left(79 z^2-59\right) \pi ^2 H_{1,1}
                -\frac{244}{27} \left(z^2+1\right) \pi ^2 H_{1,0}
\nonumber \\&
                -\frac{2}{27} \left(188 z^2+6 z+125\right) \pi ^2 H_{0,0}
                -\frac{4}{9} \left(71 z^2-47\right) \zeta _3 H_1
                -\frac{2}{9} \left(223 z^2-36 z+109\right) \zeta _3 H_0
\nonumber \\&
                +\frac{1}{405} \left(391 z^2-42 z+22\right) \pi ^4
        \bigg]
        +\frac{1}{(1-z)^2}\bigg[
                -\frac{2}{81} \left(8456 z^3-12953 z^2+10077 z-5634\right) H_{0,0}
        \bigg]
,
\nonumber\\
\intertext{and}
F_{3}(z) ={}&
        \frac{1}{2916}(715565 z-197242)
        +\frac{35}{108} (698 z-69) H_1
        +\frac{181}{27} (31 z+1) H_{1,1}
        +\frac{1}{9} (1403 z+662) H_{1,1,1}
\nonumber \\&
        +\frac{8}{3} (32 z+23) H_{1,1,1,1}
        +60 (z+1) H_{1,1,1,1,1}
        +\frac{1}{1-z}\bigg[
                \frac{1}{648}\left(-217440 z^2+191022 z-186085\right) H_0
        \bigg]
\nonumber \\&
        +\frac{1}{1-z}\bigg[
                \frac{1}{162} \left(-52174 z^2+38784 z-38101\right) H_2
                +\frac{1}{162} \left(-32914 z^2+29415 z-33625\right) H_{1,0}
\nonumber \\&
                +\frac{1}{972} \left(50848 z^2-34734 z-1747\right) \pi ^2
        \bigg]
        +\frac{1}{1-z}\bigg[
                \frac{1}{18} \left(-4800 z^2+1759 z-3599\right) H_3
\nonumber \\&
                +\frac{1}{18} \left(-3843 z^2+2024 z-3645\right) H_{2,1}
                +\frac{1}{54} \left(-8357 z^2+3903 z-8099\right) H_{2,0}
\nonumber \\&
                +\frac{1}{9} \left(-1704 z^2+795 z-1793\right) H_{1,2}
                -\frac{2}{9} \left(554 z^2-277 z+541\right) H_{1,1,0}
\nonumber \\&
                +\frac{1}{54} \left(-7033 z^2+2574 z-7429\right) H_{1,0,0}
                -\frac{13}{27} \left(407 z^2-96 z+185\right) H_{0,0,0}
\nonumber \\&
                +\frac{1}{108} \left(4442 z^2-2067 z-243\right) \pi ^2 H_1
                +\frac{1}{108} \left(6139 z^2-2356 z+4431\right) \pi ^2 H_0
\nonumber \\&
                +\frac{1}{54} \left(15898 z^2-5313 z-10099\right) \zeta _3
        \bigg]
        +\frac{1}{1-z}\bigg[
                \frac{1}{18} \left(-3653 z^2+726 z-1559\right) H_4
\nonumber \\&
                +\frac{1}{3} \left(-572 z^2+186 z-327\right) H_{3,1}
                +\frac{1}{9} \left(-1388 z^2+477 z-656\right) H_{3,0}
\nonumber \\&
                -\frac{2}{3} \left(194 z^2-35 z+132\right) H_{2,2}
                +\frac{1}{9} \left(-1163 z^2+324 z-803\right) H_{2,1,1}
                +\frac{1}{9} \left(-941 z^2+270 z-548\right) H_{2,1,0}
\nonumber \\&
                +\frac{1}{18} \left(-1369 z^2+270 z-679\right) H_{2,0,0}
                +\frac{1}{18} \left(-1925 z^2-36 z-1757\right) H_{1,3}
\nonumber \\&
                +\frac{1}{9} \left(-1015 z^2+294 z-1027\right) H_{1,2,1}
                -\frac{5}{9} \left(143 z^2-24 z+143\right) H_{1,2,0}
                -\frac{8}{9} \left(91 z^2-9 z+82\right) H_{1,1,2}
\label{eq21}
\\&
                -\frac{2}{3} \left(99 z^2-29 z+97\right) H_{1,1,1,0}
                +\frac{1}{18} \left(-749 z^2+12 z-629\right) H_{1,1,0,0}
                +\frac{1}{18} \left(-937 z^2-432 z-793\right) H_{1,0,0,0}
\nonumber \\&
                +\frac{1}{18} \left(-2166 z^2+351 z+43\right) H_{0,0,0,0}
                +\frac{1}{18} \left(531 z^2-121 z+383\right) \pi ^2 H_2
                +\frac{1}{27} \left(511 z^2-72 z-284\right) \pi ^2 H_{1,1}
\nonumber \\&
                +\frac{1}{108} \left(2723 z^2-432 z+2555\right) \pi ^2 H_{1,0}
                +\frac{1}{108} \left(4478 z^2-960 z+2195\right) \pi ^2 H_{0,0}
\nonumber \\&
                +\frac{1}{9} \left(997 z^2-180 z-769\right) \zeta _3 H_1 
                +\frac{1}{9} \left(1690 z^2-489 z+745\right) \zeta _3 H_0 
                +\frac{1}{3240} \left(-9277 z^2+6318 z-2287\right) \pi ^4
        \bigg]
\nonumber \\&
        +\frac{1}{1-z}\bigg[
                \frac{1}{2} \left(-177 z^2-23\right) H_5
                +\frac{1}{3} \left(-387 z^2-157\right) H_{4,1}
                -\frac{2}{3} \left(147 z^2+59\right) H_{4,0}
                -\frac{4}{3} \left(77 z^2+38\right) H_{3,2}
\nonumber \\&
                -2 \left(55 z^2+31\right) H_{3,1,1}
                -\frac{4}{3} \left(65 z^2+36\right) H_{3,1,0}
                -\frac{4}{3} \left(49 z^2+26\right) H_{3,0,0}
                +\frac{1}{3} \left(-235 z^2-161\right) H_{2,3}
\nonumber \\&
                -\frac{2}{3} \left(149 z^2+101\right) H_{2,2,1}
                -\frac{8}{3} \left(27 z^2+19\right) H_{2,2,0}
                -2 \left(41 z^2+29\right) H_{2,1,2}
                -12 \left(7 z^2+5\right) H_{2,1,1,1}
\nonumber \\&
                -\frac{2}{3} \left(103 z^2+73\right) H_{2,1,1,0}
                +\frac{1}{3} \left(-143 z^2-101\right) H_{2,1,0,0}
                +\left(-29 z^2-27\right) H_{2,0,0,0}
                -\frac{226}{3} \left(z^2+1\right) H_{1,4}
\nonumber \\&
                -104 \left(z^2+1\right) H_{1,3,1}
                -86 \left(z^2+1\right) H_{1,3,0}
                -\frac{220}{3} \left(z^2+1\right) H_{1,2,2}
                -78 \left(z^2+1\right) H_{1,2,1,1}
\nonumber \\&
                -64 \left(z^2+1\right) H_{1,2,1,0}
                -46 \left(z^2+1\right) H_{1,2,0,0}
                -\frac{160}{3} \left(z^2+1\right) H_{1,1,3}
                -\frac{230}{3} \left(z^2+1\right) H_{1,1,2,1}
\nonumber \\&
                -56 \left(z^2+1\right) H_{1,1,2,0}
                -\frac{160}{3} \left(z^2+1\right) H_{1,1,1,2}
                -52 \left(z^2+1\right) H_{1,1,1,1,0}
                -\frac{86}{3} \left(z^2+1\right) H_{1,1,1,0,0}
\nonumber \\&
                -10 \left(z^2+1\right) H_{1,1,0,0,0}
                -36 \left(z^2+1\right) H_{1,0,0,0,0}
                +\frac{1}{2} \left(-81 z^2-23\right) H_{0,0,0,0,0}
                +\frac{1}{360} \left(-827 z^2-81\right) \pi ^4 H_1
\nonumber \\&
                +\frac{1}{540} \left(-1296 z^2-785\right) \pi ^4 H_0
                +\frac{2}{9} \left(101 z^2+50\right) \pi ^2 H_3
                +\frac{1}{3} \left(58 z^2+41\right) \pi ^2 H_{2,1}
                +\frac{1}{9} \left(161 z^2+97\right) \pi ^2 H_{2,0}
\nonumber \\&
                +\frac{152}{9} \left(z^2+1\right) \pi ^2 H_{1,2}
                +\frac{1}{9} \left(119 z^2-115\right) \pi ^2 H_{1,1,1}
                +\frac{128}{9} \left(z^2+1\right) \pi ^2 H_{1,1,0}
                +\frac{259}{18} \left(z^2+1\right) \pi ^2 H_{1,0,0}
\nonumber \\&
                +\frac{1}{6} \left(97 z^2+5\right) \pi ^2 H_{0,0,0}
                +\frac{1}{3} \left(269 z^2+179\right) \zeta _3 H_2
                +2 \left(37 z^2-45\right) \zeta _3 H_{1,1}
                +\frac{178}{3} \left(z^2+1\right) \zeta _3 H_{1,0}
\nonumber \\&
                +\frac{1}{6} \left(409 z^2+47\right) \zeta _3 H_{0,0}
                +\frac{\left(323-363 z^2\right) }{18}  \pi ^2 \zeta _3
                +\frac{\left(287 z^2+545\right)}{6}  \zeta _5
        \bigg]
\nonumber \\&
        +\frac{1}{(1-z)^2}\bigg[
                \frac{1}{324}\left(121336 z^3-196558 z^2+139733 z-64727\right) H_{0,0}
        \bigg]
.
\nonumber 
\end{align}
\end{widetext}

We note that the NLO, NNLO and N$^3$LO contributions to the matching coefficient ${\cal I}_{qq}$
can be found in an ancillary file attached to this submission. In addition to the
functions $F^{(1,2,3)}(z)$, also the functions $F_+^{(3,k)}(z)$ and constants $C^{(3)}_{k}$ can be found there, in a computer-readable
  form.

\section{Conclusions}
\label{sect5}

In this paper, we  presented the N$^3$LO matching coefficient for the  0-jettiness quark beam function in the large-$N_c$ large-$N_f$
approximation. 
We have compared  our results for the matching coefficient
${\cal I}_{qq}$  with the results in the  literature \cite{Billis:2019vxg} and found 
perfect agreement  for all terms that are available. The new result of this paper is the hard  contribution to the matching
coefficient ${\cal I}_{qq}$ given in Eqs.~(\ref{eq19})--(\ref{eq21}). The full
matching coefficient with soft terms and $t$-dependent plus-distributions
can be found  in an ancillary file provided with this article. 

Although our large-$N_c$ large-$N_f$ result is, perhaps, not quite suitable for phenomenology per se, we believe it is an important
milestone in the computation of beam functions through N${}^3$LO QCD. Indeed, it clearly shows 
that computations of  complete  matching coefficients  for quark and  gluon beam functions  
at N$^3$LO are  within reach. In fact, although only  planar Feynman diagrams
are needed for computations in the large-$N_c$ limit,  we already have
all  the ingredients for  gluonic final states  to go beyond this approximation. We are in the process of
computing all relevant integrals  to describe $q \to q^* + q \bar q \,(+g) $ transitions;
once these integrals are obtained, going beyond the generalized large-$N_c$ approximation will be quite straightforward. 
   
{\bf Acknowledgments}
We are grateful to Daniel Baranowsky for  supplying NNLO matching coefficients expanded to higher orders
in the dimensional regularization parameter prior to their publication.  We would like to thank Claude Duhr for
answering questions about Ref.~\cite{Duhr:2014nda} and for providing a computer-readable version of the results therein.
The research of A.B., K.M. and R.R. is partially supported by the  Deutsche Forschungsgemeinschaft (DFG, German Research Foundation) under grant 396021762 - TRR 257. The research of L.T. is  supported by the Royal Society through a Royal Society University Research Fellowship, grant number: URF\textbackslash R1\textbackslash 191125,
and in part by the ERC grant 637019 ``MathAm''. The research of C.W. is supported in part by
the BMBF project No.~05H18WOCA1.
The diagrams were drawn using \texttt{JaxoDraw} \cite{Vermaseren:1994je,Binosi:2003yf,Binosi:2008ig}.

\section*{Appendix}
In this Appendix, we present explicit intermediate formulas required to express the matching coefficient
through the partonic bare beam function. 

First, we  show how  to construct an $\overline {\rm MS}$  parton distribution function in perturbation theory.
The starting point is the Altarelli-Parisi equation, Eq.~(\ref{eq:DGLAP}),
and the perturbative expansion of the splitting functions 
\be
P_{ij}(z) = \sum_{n=0}^{\infty} \left(\frac{\alpha_s}{2\pi}\right)^n  P_{ij}^{(n)}(z).
\ee
To construct the parton distribution functions $f_{ij}$, we integrate the DGLAP equation using the evolution equation
for the strong coupling constant 
\begin{align}
&\mu^2 \frac{d}{d\mu^2} \alpha_s(\mu^2) = \beta(\alpha_s) - \epsilon \,\alpha_s(\mu^2)\,,
\\
&\beta(\alpha_s) = - \frac{\alpha_s^2}{4\pi} \beta_0 - \frac{\alpha_s^3}{(4\pi)^2} \beta_1 + \mathcal{O}(\alpha_s^4)\,,
\end{align}
with the boundary condition $f_{ij}^{(0)}(z) = \delta(1-z)$ using the following formulas for the $\beta$-functions
\be
\begin{split} 
&
\beta_0 = \frac{11}{3} C_A - \frac{4}{3} T_R N_f\,,
\\
&
\beta_1 = \frac{34}{3} C_A^2 - \left(\frac{20}{3} C_A + 4 C_F \right) T_R N_f\,.
\end{split}
\ee
 We write the result for the partonic PDFs as
\be
\begin{split}
f_{ij}^{(1)} ={}& 
- \frac{1}{\epsilon} P_{ij}^{(0)}
\,,\\
f_{ij}^{(2)} ={}&
\frac{1}{2\epsilon^2} \sum_k P_{ik}^{(0)} \underset{z}{\otimes} P_{kj}^{(0)} 
\\
& + \frac{\beta_0}{4 \epsilon^2} P_{ij}^{(0)}
- \frac{1}{2\epsilon} P_{ij}^{(1)}
\label{eq:f_2}
\,,\\
f_{ij}^{(3)} ={}&
- \frac{1}{6\epsilon^3} \sum_{k,\ell} P_{ik}^{(0)} \underset{z}{\otimes} P_{k\ell}^{(0)} \underset{z}{\otimes} P_{\ell j}^{(0)}
\\& 
- \frac{\beta_0}{4\epsilon^3} \sum_k P_{ik}^{(0)} \underset{z}{\otimes} P_{kj}^{(0)}
- \frac{\beta_0^2}{12\epsilon^3} P_{ij}^{(0)}
\\&
+ \frac{1}{3\epsilon^2} \sum_k P_{ik}^{(1)} \underset{z}{\otimes} P_{kj}^{(0)} 
+ \frac{\beta_0}{6\epsilon^2} P_{ij}^{(1)}
\\&
+ \frac{1}{6\epsilon^2} \sum_k P_{ik}^{(0)} \underset{z}{\otimes} P_{kj}^{(1)}
+ \frac{\beta_1}{12\epsilon^2} P_{ij}^{(0)}
\\&
- \frac{1}{3\epsilon} P_{ij}^{(2)}\,,
\end{split}
\ee
where the dependency of $f_{ij}$'s and $P_{ij}$'s  on $z$ has been suppressed. 

Next, we write the relations between bare $B^{{\rm b}}(t,z)$ and renormalized beam functions $B(t,z,\mu)$ at various orders in $\alpha_s$. 
Writing the relevant $\alpha_s$-expansions 
\begin{align}
B^{\rm{b}}_{ij}(t,z) &= \sum_{n=0}^{\infty} \left(\frac{\alpha_s}{4\pi}\right)^n  B^{\rm{b}\,(n)}_{ij}(t,z)\,,
\label{eq:B_bare_expansion}
\\
Z^{-1}_{i}(t,\mu) &= \sum_{n=0}^{\infty} \left(\frac{\alpha_s}{4\pi}\right)^n  Z^{-1\,(n)}_{i}(t,\mu)\,,
\label{eq:Z_inverse_expansion}
\end{align}
and using the boundary conditions 
$B^{\rm{b}\,(0)}_{ij}(t,z) = \delta_{ij} \delta(t) \delta(1-z)$ and $Z^{-1\,(0)}_{i}(t,\mu) = \delta(t)\,$
in conjunction with Eq.~(\ref{eq8}), we obtain 
\be
\begin{split}
B_{ij}^{(1)} ={}& 
B_{ij}^{\rm{b}\,(1)}
+ \delta_{ij}\delta(1-z) Z^{-1\,(1)}_{i}\,,
\\
B_{ij}^{(2)} ={}&
B_{ij}^{\rm{b}\,(2)} 
+ Z^{-1\,(1)}_{i} \underset{t}{\otimes} B_{ij}^{\rm{b}\,(1)}
\\& 
+ \delta_{ij}\delta(1-z) Z^{-1\,(2)}_{i}\,,
\\
B_{ij}^{(3)} ={}&
B_{ij}^{\rm{b}\,(3)} 
+ Z^{-1\,(1)}_{i} \underset{t}{\otimes} B_{ij}^{\rm{b}\,(2)} 
\\&
+ Z^{-1\,(2)}_{i} \underset{t}{\otimes} B_{ij}^{\rm{b}\,(1)} 
+ \delta_{ij}\delta(1-z) Z^{-1\,(3)}_{i}\,.
\label{eq:B_3}
\end{split}
\ee
The relevant renormalization coefficients for $i=q$ read 
\begin{align}
\MoveEqLeft{Z^{-1\,(1)}_{q} =
        C_F \left[
                \frac{4 }{\epsilon }  L_0\left (\frac{t}{\mu ^2}\right)
                -\delta (t)\left(\frac{4}{\epsilon ^2}+\frac{3}{\epsilon }\right)
        \right]
}
\,, \\
\MoveEqLeft{Z^{-1\,(2)}_{q} =
        C_F^2 \bigg[
                \frac{16 }{ \epsilon ^2}  L_1\left(\frac{t}{\mu ^2}\right)
                -L_0\left(\frac{t}{\mu ^2}\right) \left(
                        \frac{16}{\epsilon ^3}
                        +\frac{12}{\epsilon ^2}
                \right) 
}&\nonumber\\&
                +\delta(t) \biggl(
                        \frac{8}{\epsilon ^4}
                        +\frac{12}{\epsilon ^3}
                        +\frac{1}{\epsilon ^2}\left(\frac{9}{2}-\frac{4 \pi ^2}{3}\right)
                        +\frac{1}{\epsilon }\left(
                                -\frac{3}{4}
\right.\nonumber\\&\left.
                                +\pi ^2
                                -12 \zeta_3
                        \right)
                \biggr)
        \bigg]
        +C_A C_F \bigg[
                L_0 \left ( \frac{t}{\mu ^2}\right) \biggl(
                        -\frac{22}{3 \epsilon ^2}
\nonumber\\&
                        +\frac{1}{\epsilon }\left(\frac{134}{9}-\frac{2 \pi ^2}{3}\right)
                \biggr)
                +\delta (t) \biggl(
                        \frac{11}{\epsilon ^3}
                        +\frac{1}{\epsilon ^2} \left(
                                -\frac{35}{18}
\right.\\&\left.
                                +\frac{\pi ^2}{3}
                        \right)
                        +\frac{1}{\epsilon }\left(-\frac{1769}{108}-\frac{11 \pi ^2}{18}+20 \zeta_3\right)
                \biggr)
        \bigg]
\nonumber\\&
        +C_F N_f T_F \bigg[
                L_0 \left (\frac{t}{\mu ^2}\right) \left(
                        \frac{8}{3 \epsilon ^2}
                        -\frac{40}{9 \epsilon }
                \right)
\nonumber \\ &
                +\delta (t) \left(
                        -\frac{4}{\epsilon ^3}
                        +\frac{2}{9 \epsilon ^2}
                        +\frac{1}{\epsilon }\left(\frac{121}{27}+\frac{2 \pi ^2}{9}\right)
                \right) 
        \bigg]
\,, \nonumber 
\end{align}
and
\begin{widetext}
\begin{align}
Z^{-1\,(3)}_{q} ={}&
        C_F^3 \bigg[
                \frac{32}{\epsilon ^3}  L_2\left ( \frac{t}{\mu ^2}\right)
                -\left(\frac{64}{\epsilon ^4}+\frac{48}{\epsilon ^3}\right)  L_1 \left ( \frac{t}{\mu ^2}\right)
                + L_0 \left ( \frac{t}{\mu ^2}\right) \biggl(
                        \frac{32}{\epsilon ^5}
                        +\frac{48}{\epsilon ^4}
                        +\frac{1}{\epsilon ^3}\left(18-\frac{16 \pi ^2}{3}\right)
\nonumber\\&
                        +\frac{1}{\epsilon ^2}\left(-3+4 \pi ^2-48 \zeta _3\right)
                \biggr)
                +\delta (t) \bigg(
                        -\frac{32}{3 \epsilon ^6}
                        -\frac{24}{\epsilon ^5}
                        +\frac{1}{\epsilon ^4}\left(-18+\frac{16 \pi ^2}{3}\right)
                        +\frac{1}{\epsilon ^3}\left(-\frac{3}{2}+\frac{208 \zeta _3}{3}\right)
\nonumber\\&
                        +\frac{1}{\epsilon ^2}\left(\frac{9}{4}-3 \pi ^2+36 \zeta _3\right)
                        +\frac{1}{\epsilon}\left(
                                -\frac{29}{6}
                                -\pi ^2
                                -\frac{68 \zeta _3}{3}
                                -\frac{8 \pi ^4}{15}
                                +\frac{16 \pi ^2 \zeta _3}{9}
                                +80 \zeta _5
                        \right)
                \bigg) 
        \bigg]
\nonumber\\&
        +C_A C_F^2 \bigg[
                L_1 \left ( \frac{t}{\mu ^2}\right) \left(
                        -\frac{176}{3 \epsilon ^3}
                        +\frac{1}{\epsilon ^2}\left(\frac{1072}{9}-\frac{16 \pi ^2}{3}\right)
                \right) 
                + L_0 \left ( \frac{t}{\mu ^2}\right) \biggl(
                        \frac{220}{3 \epsilon ^4}
                        +\frac{1}{\epsilon ^3}\left(-\frac{136}{3}+4 \pi ^2\right)
\nonumber\\&
                        +\frac{1}{\epsilon ^2}\left(-\frac{2975}{27}-\frac{4 \pi ^2}{9}+80 \zeta _3\right)
                \biggr)
                +\delta (t)\bigg(
                        -\frac{44}{\epsilon ^5}
                        -\frac{1}{\epsilon ^4}\left(\frac{227}{9}+\frac{4 \pi ^2}{3}\right)
                        +\frac{1}{\epsilon ^3}\left(\frac{3853}{54}+\frac{19 \pi ^2}{3}-80 \zeta _3\right)
\nonumber\\&
                        +\frac{1}{\epsilon ^2}\left(\frac{1835}{36}-\frac{569 \pi ^2}{54}-\frac{92 \zeta _3}{3}+\frac{4 \pi ^4}{9}\right)
                        +\frac{1}{\epsilon }\left(-\frac{151}{12}+\frac{205 \pi ^2}{27}-\frac{844 \zeta _3}{9}+\frac{247 \pi ^4}{405}-\frac{8}{9} \pi ^2 \zeta _3-40 \zeta _5\right)
                \bigg) 
        \bigg] 
\nonumber\\&
        + C_A^2 C_F \bigg[
                L_0 \left ( \frac{t}{\mu ^2}\right) \left(
                        \frac{484}{27 \epsilon ^3}
                        +\frac{1}{\epsilon ^2}\left(-\frac{4172}{81}+\frac{44 \pi ^2}{27}\right)
                        +\frac{1}{\epsilon }\left(\frac{490}{9}-\frac{536 \pi ^2}{81}+\frac{88 \zeta _3}{9}+\frac{44 \pi ^4}{135}\right)
                \right)
\nonumber\\&
                +\delta (t) \bigg(
                        -\frac{2662}{81 \epsilon ^4}
                        +\frac{1}{\epsilon ^3}\left(\frac{8999}{243}-\frac{110 \pi ^2}{81}\right)
                        +\frac{1}{\epsilon ^2}\left(\frac{16147}{486}+\frac{899 \pi ^2}{243}-\frac{1408 \zeta _3}{27}-\frac{44 \pi ^4}{405}\right)
\nonumber\\&
                        +\frac{1}{\epsilon }\left(-\frac{412907}{8748}-\frac{419 \pi ^2}{729}+\frac{5500 \zeta _3}{27}-\frac{19 \pi ^4}{30}-\frac{88}{27} \pi ^2 \zeta _3-\frac{232 \zeta _5}{3}\right)
                \bigg)
        \bigg]
\\&
        +C_F^2 N_f T_R \bigg[
                L_1 \left ( \frac{t}{\mu ^2}\right) \left(\frac{64}{3 \epsilon ^3}-\frac{320}{9 \epsilon ^2}\right) 
                + L_0\left(\frac{t}{\mu ^2}\right) \biggl(
                        -\frac{80}{3 \epsilon ^4}
                        +\frac{32}{3 \epsilon ^3}
                        +\frac{1}{\epsilon ^2}\left(\frac{988}{27}+\frac{8 \pi ^2}{9}\right)
\nonumber\\&
                        +\frac{1}{\epsilon }\left(-\frac{220}{9}+\frac{64 \zeta _3}{3}\right)
                \biggr)
                +\delta (t) \bigg(
                        \frac{16}{\epsilon ^5}
                        +\frac{100}{9 \epsilon ^4}
                        -\frac{1}{\epsilon ^3}\left(\frac{694}{27}+\frac{8 \pi ^2}{3}\right)
                        +\frac{1}{\epsilon ^2}\left(-\frac{269}{27}+\frac{86 \pi ^2}{27}-\frac{160 \zeta _3}{9}\right)
\nonumber\\&
                        +\frac{1}{\epsilon }\left(\frac{4664}{81}-\frac{32 \pi ^2}{27}+\frac{208 \zeta _3}{27}-\frac{164 \pi ^4}{405}\right)
                \bigg) 
        \bigg]
\nonumber\\&
        + C_A C_F N_f T_R \bigg[
                L_0 \left ( \frac{t}{\mu ^2}\right) \left(
                        -\frac{352}{27 \epsilon ^3}
                        +\frac{1}{\epsilon ^2}\left(\frac{2672}{81}-\frac{16 \pi ^2}{27}\right)
                        +\frac{1}{\epsilon }\left(-\frac{1672}{81}+\frac{160 \pi ^2}{81}-\frac{224 \zeta _3}{9}\right)
                \right)
\nonumber\\&
                +\delta (t) \bigg(
                        \frac{1936}{81 \epsilon ^4}
                        +\frac{1}{\epsilon ^3}\left(-\frac{5384}{243}+\frac{40 \pi ^2}{81}\right)
                        +\frac{1}{\epsilon ^2}\left(-\frac{6148}{243}-\frac{424 \pi ^2}{243}+\frac{704 \zeta _3}{27}\right)
\nonumber\\&
                        +\frac{1}{\epsilon }\left(-\frac{5476}{2187}+\frac{1180 \pi ^2}{729}-\frac{2656 \zeta _3}{81}+\frac{46 \pi ^4}{135}\right)
                \bigg) 
        \bigg]
\nonumber\\&
        +C_F N_f^2 T_R^2 \bigg[
                L_0 \left (\frac{t}{\mu ^2}\right) \left(
                        \frac{64}{27 \epsilon ^3}
                        -\frac{320}{81 \epsilon ^2}
                        -\frac{64}{81 \epsilon }
                \right) 
                +\delta (t)\bigg(
                        -\frac{352}{81 \epsilon ^4}
                        +\frac{368}{243 \epsilon ^3}
                        +\frac{1}{\epsilon ^2}\left(\frac{344}{81}+\frac{16 \pi ^2}{81}\right)
\nonumber\\&
                        +\frac{1}{\epsilon }\left(\frac{13828}{2187}-\frac{80 \pi ^2}{243}-\frac{256 \zeta _3}{81}\right)
                \bigg) 
        \bigg]
\,. \nonumber
\end{align}
\end{widetext}  

The NLO and NNLO coefficients agree with Ref.~\cite{Ritzmann:2014mka}.

\bibliographystyle{apsrev4-1}
\bibliography{bibliography}

\end{document}